Corresponding Author: Mrs. VALERIA MANGANO, Ph.D.

Corresponding Author's Institution: INAF- Istituto Nazionale di Astrofisica

First Author: VALERIA MANGANO, Ph.D.

Order of Authors: VALERIA MANGANO, Ph.D.; STEFANO MASSETTI; ANNA MILILLO; ALESSANDRO MURA; STEFANO ORSINI; FRANCOIS LEBLANC

Abstract: The exosphere, the tenuous collisionless cloud of gas surrounding Mercury is still a poorly known object because it is the result of many various interactions between the surface, the interplanetary medium (Solar wind, photons and meteoroids), the planetary and the interplanetary magnetic fields. Many ground-based observations have allowed the detection of intense and variable sodium emissions at global and local spatial scales, the latter being mostly concentrated in the polar-mid latitude regions. These regions are indeed the preferred location of solar wind precipitation on the surface of the planet.
In the present paper, by using high resolution Na observations obtained at the Canary Islands with the THEMIS solar telescope, we analyze the variability of the sodium exosphere on time-scale of 1 hour and investigate the possible mechanisms that could explain the exospheric sodium emission distribution and its dynamics. Our interpretation relates the observed sodium asymmetries to the combined effects of plasma and photons impacts onto the Mercury's surface and of sodium diffusion through the upper layer of the surface. The comparison between data and simulations seems to evidence that, similarly to what occurs at the Earth, both the magnetic reconnection regimes of pulsed or quasi-steady reconnection could occur on Mercury, and be responsible for the observed Na short term variations. In addition to this, a progressive broadening of the peak regions together with an increase of the equatorial region seem to corroborate the idea of the role of photon stimulated desorption, in association with ion sputtering and with global sodium migration around Mercury as the cause of the observed evolution of the Na exosphere.

**HIGHLIGHTS:**

- Observations of sodium emission from the exosphere of Mercury are performed.
- The analysis evidences the well-known two peaks feature, and a variability on the time-scale of 1 hour.
- A devoted procedure is used to allow comparison among the different data of the day (each with a different seeing value).
- Models of exosphere and magnetosphere are used to explain the observed features.

# DYNAMICAL EVOLUTION OF SODIUM ANYSOTROPIES

# IN THE EXOSPHERE OF MERCURY

Valeria Mangano (corresponding author)

INAF/IAPS, via del Fosso del Cavaliere, 100 - 00133 Roma, Italy

Tel.: +39 06 45488752; Fax: +39 06 4993-4242

valeria.mangano@iaps.inaf.it

Stefano Massetti

INAF/IAPS, via del Fosso del Cavaliere, 100 - 00133 Roma, Italy

stefano.massetti@iaps.inaf.it

Anna Milillo

INAF/IAPS, via del Fosso del Cavaliere, 100 - 00133 Roma, Italy

anna.milillo@iaps.inaf.it

Alessandro Mura

INAF/IAPS, via del Fosso del Cavaliere, 100 - 00133 Roma, Italy

alessandro.mura@iaps.inaf.it

Stefano Orsini

INAF/IAPS, via del Fosso del Cavaliere, 100 - 00133 Roma, Italy

stefano.orsini@iaps.inaf.it

Francois Leblanc

LATMOS/IPSL, Université Versailles Saint Quentin, CNRS, France

francois.leblanc@latmos.ipsl.fr

**ABSTRACT**

The exosphere, the tenuous collisionless cloud of gas surrounding Mercury is still a poorly known object because it is the result of many various interactions between the surface, the interplanetary medium (Solar wind, photons and meteoroids), the planetary and the interplanetary magnetic fields. Many ground-based observations have allowed the detection of intense and variable sodium emissions at global and local spatial scales, the latter being mostly concentrated in the polar-mid latitude regions. These regions are indeed the preferred location of solar wind precipitation on the surface of the planet.

In the present paper, by using high resolution Na observations obtained at the Canary Islands with the THEMIS solar telescope, we analyze the variability of the sodium exosphere on time-scale of 1 hour and investigate the possible mechanisms that could explain the exospheric sodium emission distribution and its dynamics. Our interpretation relates the observed sodium asymmetries to the combined effects of plasma and photons impacts onto the Mercury's surface and of sodium diffusion through the upper layer of the surface. The comparison between data and simulations seems to evidence that, similarly to what occurs at the Earth, both the magnetic reconnection regimes of pulsed or quasi-steady reconnection could occur on Mercury, and be responsible for the observed Na short term variations. In addition to this, a progressive broadening of the peak regions together with an increase of the equatorial region seem to corroborate the idea of the role of photon stimulated desorption, in association with ion sputtering and with global sodium migration around Mercury as the cause of the observed evolution of the Na exosphere.

# 1 - INTRODUCTION

The exosphere of Mercury is a complex object whose composition and dynamics are the result of the diverse interactions that occur between the Hermean surface and magnetic field on one side, and the interplanetary medium on the other. In fact, solar wind particles and photons, meteoritic impacts and interplanetary magnetic field have a role on the morphology, content and time evolution of this tiny gaseous envelope (e.g.: Killen and Ip, 1999; Milillo et al. 2005).

Since the discovery of the sodium component in the Hermean exosphere in 1985 (Potter and Morgan, 1985), many observations evidenced a significant temporal variability of the average sodium intensity (Potter et al. 2006; Leblanc and Johnson 2010) and of high latitudes localized peaks of emissions (e.g.: Potter and Morgan, 1990; 1997a; Killen *et al.*, 1990; Potter *et al.*, 1999). Such peaks may appear at mid latitude in both hemispheres, or be localized only in one hemisphere, or may display a maximum intensity close to the equator. To explain such behavior, a correlation with the IMF has been widely supposed (e.g.: Massetti *et al.* 2003, 2007, Kallio *et al.* 2003, Sarantos *et al.* 2001), since its orientation is expected to allow preferential precipitation of solar wind protons into one cusp region or both, by driving the topology of the magnetic reconnection with the internal magnetic field. The direct proton impacts on the surface would then cause proportional Na release via processes related to these impacts, like the ion-sputtering process. Furthermore, the observed short term variability cannot be easily explained by any other known mechanisms (Leblanc et al. 2009). Nevertheless, due to the estimated low efficiency of atom release, the ion-sputtering (IS) process alone cannot account for the observed column density: in fact, the Na density in the exosphere due to IS was estimated to be about 2 orders of magnitude less intense than actually measured by ground based observations (Mura et al., 2007, Wurz et al., 2010). On the contrary, the intensity of sodium atoms released by the photon-stimulated desorption (PSD) is in agreement with the observed intensities (Madey et al., 1998; Killen and Ip, 1999) but cannot account for the frequent asymmetries observed in exospheric emission, unless they are due to surface sodium concentration peaks (Sprague and Massey, 2007 Leblanc and Johnson, 2003). Many investigations have been performed trying to relate peak of emission and surface features, nevertheless statistical analysis (Potter et al.,2006) did not show any relation of sodium intensity with surface longitude.

Also the idea of a major meteoroid impact producing a dense cloud due to surface vaporization is not very likely, since this process does not match morphologically with the evidence of the observations, often showing peaks at mid-latitudes on both hemispheres (with variable relative intensity). Such an event should be expected to act not only at mid-latitudes but on the whole disk of Mercury with the same probability. Moreover, as estimated by Mangano et al. (2007), the frequency of meteoroid impacts with size compatible to the ground-based observations Na intensity is much lower than what is usually observed.

Mura et al. (2009) compared the sodium exosphere observations made by Schleicher et al. (2004) to the result of a numerical simulation. The observations made during Mercury transit above the solar disk show the Na exosphere to be denser at high latitudes and show a clear dawn-dusk asymmetry. To explain the latitudinal asymmetries and the intensities, Mura et al. (2009) suggested that the Na exosphere generation involves two steps. The first step consists of the solar wind proton precipitation causing chemical alteration of the surface, freeing the sodium atoms from their bounds in the crystalline structure on the surface. The second step is photon-stimulated desorption (PSD) and thermal desorption (TD) of this Na into the exosphere. The slow rotation of the planet would permit a progressive sedimentation of Na atoms from the exosphere on the night-side surface, thus providing a sodium reservoir as originally suggested by Hunten and Sprague (1997) and modeled by Leblanc and Johnson (2003). In the first hours of dawn the Na is then released again from the surface due to PSD process. The good agreement of the simulated exosphere with the Schleicher et al. (2004) observations supported the proposed scenario. Similar mechanisms of combined action of proton precipitation and photon impact on the surface have been proposed also by Wilson et al. (2006). They studied the Na exosphere of the Moon during the passage in the Earth's magnetotail and magnetosheath. The increase of the Na emission intensity when the Moon enters in the denser magnetosheath was explained by an enhancement of the diffusion of the Na atoms through the regolith grains induced by ion impacts (Potter et al., 2000), inducing an higher PSD release.

From sparse observations, it is of course very difficult to investigate the dynamics of Na exosphere. Recently, the possibility to perform observations from the Solar telescope THEMIS brought the opportunity to have exospheric observations at both high spatial and time resolution, including measurements of the radial velocity component through the sodium line Doppler shift. In this way, it has been possible to investigate both intensity and velocity variations (Leblanc et al., 2008). Leblanc et al. (2008) found a relation of high latitude intensifications and increase of the radial velocity, suggesting that the ejection mechanism during these events is significantly more energetic than the mechanism(s) producing the bulk of the exosphere. They explained these features with PSD process associated to local maxima of sodium surface density (Leblanc et al., 2007) or ion sputtering. In a second set of observations, Leblanc et al. (2009) observed a short time scale variation of the sodium exospheric emission spatial intensity and concluded as being a strong argument in favor of ion-sputtering induced by solar wind precipitation.

In this paper we analyze a long time set of observations obtained by means of THEMIS solar telescope, already partly presented in Leblanc et al. (2009), to investigate the variability of both intensity and morphology of the high latitude intensifications. In Section 2, we describe the dataset of Mercury's sodium emission recorded between July $12^{th}$-$18^{th}$ 2008, and analyze the Na variability in the time scale of 1-hour for one special day (July $13^{th}$) when remarkably favorable seeing conditions allowed the record of a long series of good quality data.

The model by Mura et al. (2008) used to simulate data and an updated version of the magnetospheric model by Massetti et al. (2007) are briefly described in Section 3, together with their application to the present case.

In Section 4, the results are discussed and compared to the expectations of the two models in order to investigate the possible mechanisms of sodium release and dynamics that may explain the observed behavior. Conclusions are summarized in Section 5.

## 2 - OBSERVATIONS

THEMIS (Lòpez Ariste et al., 2000) is a solar telescope with a 0.9 m primary mirror and a 15.04 m focal length, and is located at 2400 meters a.s.l. on the Teide volcano in Tenerife (Canary Islands). Since 2007, THEMIS is used for a long term campaign of observations of the Mercury exosphere (Leblanc et al., 2008; Leblanc et al., 2009; Leblanc and Johnson, 2010; Leblanc et al. 2012). The use of a solar telescope allows day-long observations and high resolution imaging with a resolving power of at least 220000. Moreover, two individual cameras are used to image at the same time the spectral regions around the D1 at 5896 Å and D2 at 5890 Å Na emission lines. In case of good sky conditions, high spectral resolution can be used with resolving power up to 440000. Thanks to the use of a long slit scanning the disk of Mercury together with tip-tilt corrections at ~1 kHz, good quality images of the Mercury sodium exosphere can be obtained,.

In July 2008, Mercury was imaged between July 12$^{th}$ and July 18$^{th}$, when the true anomaly angle (TAA) was in the range 303°-341°, the phase angle was increasing from -70° to -45° and the illuminated portion of the disk (showing the dusk terminator) was increasing from 66% to 85%. The technique of data acquisition and reduction is the same used since the first observations in 2007. Data are bias corrected and flat fielded. Spectral calibration is performed by using telluric lines on a sky spectrum imaged every day during the observations. The sky background at the sodium D lines wavelength is obtained by interpolating the values at the two ends of the long slit. The exospheric emission lines are finally integrated after subtracting an average background level. Brightness calibration is done following the method developed by Sprague et al. (1997) using Hapke model to derive the expected surface brightness.

In July 2008 a whole week of good and stable weather allowed high quality and continuous measurements, resulting in a remarkable sequence of bi-dimensional images of the sodium emission from the Hermean exosphere. In particular, the second day of the run (July 13$^{th}$, 2008) appears to be the most interesting; this was already evidenced in Leblanc et al. (2009) where the first high resolution data obtained at THEMIS were presented. In that paper, six consecutive images of the brightness, Doppler shift and width of Mercury's exospheric D2 emission line have been obtained during almost eleven hours of observation at the THEMIS solar telescope. The paper by Leblanc et al. (2009) was mostly devoted to the study of energy and thermal spread of the exospheric sodium cloud, but it did not analyze in detail the dynamical evolution of the two-peak sodium pattern. This is what we will do in the following, also by applying a very careful method to allow proper inter-comparison among the data (to be described later in this section).

**TABLE 1**

**FIGURE 1**

On July 13th, very clear sky condition allowed the collection of a series of high resolution data (except the last two acquisitions, collected at low resolution). The seeing, in fact, was very stable all day long, and always better than 2 arcsecs. Table 1 summarizes the details of the selected scans in July 13th that we will use for our analysis, and Figure 1 shows the bi-dimensional maps obtained after data reduction.

**FIGURE 2**

In Figure 2 we plot the average intensity emission (in kR) versus time of each scan performed during the six days under analysis. The average values are computed by taking into account only the emission coming from the left-side half disk of Mercury, within 1.2 Hermean radii ($R_M$ from now on) from its center; with this choice we guarantee to include all the main features of the Na exosphere visible in the illuminated side of the planet. The figure puts in evidence a clear decreasing trend during the whole period which matches the known dependence of the mean emission of the exosphere of Mercury with respect to true anomaly angle (TAA), as firstly evidenced by Smyth and Marconi (1995), and further confirmed by Potter et al. (2007).

In order to properly compare the Na emission from these different scans, we remove the TAA dependence by normalizing the mean emission values plotted in Figure 2 to the fit of the average trend. Secondly, to take into account possible effects caused by changes in the seeing (see Table 1), we degrade all the images to match the reference seeing value of 1.77" (scan 12). Hence, for this analysis we reject all the scans with seeing > 1.77". The degradation is achieved by making a convolution between the image and a normalized Gaussian function with a proper FWHM, so that the final FWHM value of the equivalent Gaussian is exactly 1.77". This procedure is possible since for long exposures (i.e., >> 100 ms) the point spread function is well modeled by a Gaussian, and by assuming iso-planatism, which allows us to consider the single 'patch' of 4-5 arcsec (hence, the whole disk of Mercury) as affected by one single value of seeing.

**FIGURE 3**

The resulting final six scans are displayed in Figure 3. Here a major emission peak in the southern hemisphere and a minor peak in the Northern hemisphere are clearly visible at the beginning, as well as a progressive decrease of their intensities during the day. In the second observation (Figures 1*b*) we notice a secondary intensification of the Na emission occurring equatorward of the southern main peak. This intensification is blurred but still visible in the modified image of Figure 3*b*.

# 3- MODELS FOR DATA INTERPRETATION

## 3.1 Exospheric model

The exosphere model used here (Mura et al. 2008; Mura, 2011) is able to simulate the time evolution of the sodium exosphere and surface concentration, so that, given a sodium source at a definite time and location, it is possible to estimate its concentration in any other point of the surface or of the exosphere at a subsequent time. The model assumes a concentration of sodium on the surface variable both in space and time. In fact, the topmost layer has a limited availability of sodium, so that both photon-stimulated desorption (PSD) and thermal desorption (TD) can produce a limited flux of this species, and tend to decrease its surface concentration during the Hermean day. Diffusion of sodium from the inside tend to supply the surface concentration; in addition to that, proton precipitation can efficiently accelerate the replenishment of "fresh" sodium in the uppermost layers of the surface ("chemical sputtering" or "enhanced diffusion", see Mura et al., 2008, Sarantos et al., 2008). In summary, the concentration (and, hence, the emission) of sodium from the surface of Mercury depends on the plasma precipitation, either through ion or chemical sputtering or enhanced diffusion, or all of them together. However, the plasma precipitation at the time of the observation is not known, neither it is possible to simulate it because the solar wind IMF is unknown as well, so that it is necessary to do some assumptions.

In the Na observation obtained by occultation technique by Schleicher et al. (2004) the major sodium emission occurs in the northern hemisphere. In order to reproduce the observed exosphere, Mura et al. (2008) considered a specific IMF condition with a IMF $B_x < 0$, interacting with a dipolar internal planet-centered magnetic field. Preliminary simulations show that, under similar IMF conditions, integrating the column density along the proper viewing angle and with a reversed and smaller (~50%) x component of the IMF, the morphology of the peaks observed in July 13$^{th}$ 2008 can be reproduced in a satisfactory way (state "A"). In addition to that, to better reproduce also the equatorial intensity, we include a steady diffusion term as discussed in Mura et al. (2009). Here this term is set to $5 \cdot 10^6$ cm$^{-2}$ s$^{-1}$, in agreement with the results by Killen et al., 2004 (state "B").

Finally, to reproduce a realistic condition where the source varies in a time scale of hours, we assume that the sodium source gradually and linearly changes from a global state "A+B" at time 06:52 UT, to a situation of "B" at time 08:16 UT. The resulting outputs, giving the generated exosphere at the time of the first two scans and shown in Figure 4 (panels *a* and *b*), are in good agreement with observations, as expected. Starting from this initial configuration, this time evolving model simulates the exospheric contents as a function of time in the subsequent period. Results are shown in Figure 4 (panels *c* to *f*).

**FIGURE 4**

## 3.2 Magnetospheric model

New measurements achieved by the Messenger spacecraft now orbiting around Mercury are currently improving our knowledge about the magnetosphere of the planet, confirming the idea that it possesses an Earth-like magnetosphere generated by an intrinsic magnetic dipole. Anderson et al. (2011) recently reported a data fit giving a dipole strength of 195±10 nT·$R_M^3$ (about one third lower than previously supposed), which also seems to be (unexpectedly) shifted northward by about 0.2 $R_M$. On the base of these new findings from Messenger, we updated the Massetti et al. (2007) magnetospheric model of Mercury (see Figure 5 *a*), which was derived from ad hoc modification of the original Toffoletto-Hill (1993) analytical model, originally developed to mimic the Earth's magnetosphere. The TH93 model have the distinctive characteristics to allow the magnetic reconnection between the inner dipole and all the three IMF components, including the IMF $B_x$ (radial) component, which is expected to dominate at the orbit of Mercury.

We now assume that the distribution of the two-peak-sodium emission is driven by the magnetic reconnection between the inner magnetic dipole field and the interplanetary magnetic field through the proton entrance and precipitation toward the surface. The Hermean magnetosphere is expected to be largely controlled by both the IMF $B_x$ (radial) and $B_z$ (vertical) components (Massetti et al., 2007). The radial component is statistically the strongest and more stable IMF component at the orbit of Mercury (see Fig. 1 in Sarantos et al. 2007), and is foreseen to cause a strong asymmetry in the magnetic reconnection geometry, with broader Southern/Northern cusp during periods with positive/negative IMF $B_x$. In this context, the brightest southern Na emission we observe would suggest the IMF $B_x$ (radial) component to be positive during the whole period under analysis. But this North-South asymmetry can be enhanced toward the South hemisphere if the dipole moment of the planet is shifted northward by 0.2 $R_M$ as proposed by Anderson et al. (2011), which inevitably causes a broadening of the southern cusp footprint. Hence, by taking into account the northward shift of the inner dipole, we find that a first-order match between our magnetospheric model and the first scan can be obtained by choosing a negative IMF $B_x$, namely IMF = -10, 0, -30 nT. Figure 5 *b* shows the superposition of the modeled cusp footprints and the Na emission peaks observed at 06:52 UT (first scan). The two regions agree quite well with each other.

**FIGURE 5**

**4 - DISCUSSION**

**4.1 Data Analysis**

As already explained in Section 2, Figure 3 shows the observed Na exosphere on July 13$^{th}$ 2008, normalized in order to limit observational effects on the spatial distribution of the emission of one scan with respect to another. From the evolution of the intensity maps, we can study the evolution of the exospheric distribution during that sequence of observations. The typical two-peaks pattern is clearly visible from the first scan, and this feature remains for the whole period. The northern peak seems to fade away at the fourth scan, while the southern one remains remarkably enhanced until the end of our observations.

**FIGURE 6**

We then divide the disk of Mercury, within 1.2 $R_M$ from the center of the disk, into three different latitudinal regions named 'N', 'E' and 'S', identified with $Z > 1/3\ R_M$, $-1/3\ R_M < Z < 1/3\ R_M$ and $Z < -1/3\ R_M$, respectively. We plot the average intensity values measured within each of these regions in Figure 6. Such plot highlights the macroscopic behavior of the three different regions, and in particular the similar general decrease of S and N mean intensity, while E mean intensity increases. A second remarkable increase of intensity in the N region during the last scan will not be taken into account in this discussion because the change to 'low resolution' mode may be responsible of this difference. Finally, the southern region S appears to be noticeably brighter in intensity with respect to the other two regions during all the period of analysis.

In order to detail the spatial distribution of the emission, we further divide the disk of Mercury into 0.1 $R_M$-thick slices in latitude (i.e., perpendicular to the Z axis of Figure 3). The resulting intensity profiles are plotted in Figure 7.

**FIGURE 7**

In the upper panel, a uniform exospheric bulk emission is superposed (solid black line). The 'bulk exosphere' is taken from the exospheric model (Section 3.1) and is normalized to the equatorial intensity of the first observation. It is expected to reproduce the exospheric density distribution when the mid-latitude intensifications are absent. As a consequence, the subtraction of this simulated 'bulk' exosphere intensity would give an estimation of the net exospheric emission related to the mid-latitudes enhancements. In the middle and bottom panels of Figure 7 the resulting emission profiles for the six selected scans, i.e. after subtraction of the bulk exosphere are shown (time increases from bottom to top). In these two plots the variable morphology of the peaks (in both amplitude and position) is highlighted. More specifically, it can be noticed that both peaks in the northern and southern regions of the first three observations gradually decrease with time. These peaks seem to invert their decreasing trend during the fourth scan (10:50-12:02 UT); and then to slightly move poleward (see the right shifts of the north peak), while the equatorial region intensity continues to increase.

**4.2 Data Interpretation through the models**

We try to interpret the data in the frame of two different possible scenarii, that is, taking into account two possible magnetic reconnection regimes: pulsed reconnection (i.e. per-event based) or (quasi-)steady reconnection. Both regimes have been observed to take place on the dayside magnetopause of the Earth's magnetosphere (see for example, Phan et al., 2005) and they may be expected to occur at Mercury as well. In fact, flux transfer events (FTEs), which are the marks of pulsed magnetic reconnections, have been recently identified also in the magnetosphere of Mercury (Slavin et al., 2010). Conversely, persistent proton cusp aurora under stable and favorable IMF conditions, as observed for example by the IMAGE satellite (Frey et al., 2002), is an observational signature of the (quasi-)steady magnetic reconnection regime, which can be likely put in relation with the persistent cusp-related Na emission frequently detected in the exosphere of Mercury from ground-based observations.

Due to the lack of in-situ information about the Solar Wind and IMF configuration at the time of our observation, it's awkward for us to discriminate between the two reconnection regimes in the present case study. Nevertheless, we can try to depict a realistic scenario able to explain the observed features evolution.

The observations sequence seems to indicate the occurring of ion precipitation along the open field lines through the polar cusps. If we adopt the scenario of pulsed magnetic reconnection, a first precipitation event could have occurred at the beginning of the period (first scan) and then *at least* a second one (of lower intensity effects) during the fourth scan. This second event seems to be associated with a poleward movement of the peak positions, especially in the northern hemisphere, even if the seeing uncertainty does not allow us to state it without any doubt. This could be the signature of variation in the IMF conditions. In fact, the increase of IMF $B_z$ component would, indeed, cause a poleward movement of the cusps. In this frame, we can compare our data with the exospheric model that assumes a single precipitation (i.e. magnetic reconnection) event responsible for the first observed southern peak (first scan). Trying to fit the whole observation sequence trend including the secondary peak at the fourth scan by applying *ad hoc* variable parameters would be possible, but not very significant for the interpretation. The comparison between data (Figure 3) and simulations (Figure 4) gives accordance in the column density magnitude (first scan maximum values of $1 \cdot 10^{10}$ vs $1.2 \cdot 10^{10}$ at/cm$^2$), while the modeled Na density shows a decrease rate of both the northern and southern peak intensities that is faster than the observed one. In fact, the simulated event intensification is almost finished in the fourth scan (10:50-12:02 UT), that is in few hours, in accordance with the time scale estimated for this process at TAA = 300° in Mura (2011) between 2-5 hours. The decay time derived from the observations, instead, is remarkably longer. This can be likely due to the fact that the first ion precipitation event wasn't actually a prompt ion injection as assumed, but a peak in the ion precipitation flux which continued for a period of time longer than the typical decay time-scale. On the other hand, the comparison between data and simulations accounts for the fact that the model seems to describe properly two other aspects evidenced in Figure 7, which are: 1. the width of the Na emission that is progressively broadening; 2. the migration of the sodium away from the place of extraction that is producing a progressive increase of the equatorial intensity level. Actually, the exospheric model describes the creation and subsequent migration of the exospheric sodium after the first ion precipitation event and such a behavior seems to really occur in the data.

Conversely, if we look at the ion precipitation along the cusps as a (quasi-)continuous flow caused by (quasi-)steady magnetic reconnection, then it is a more complex task to try to decouple the continuous "fresh" Na emission from the migration of "older" Na (i.e. previously released). Since the two effects can overlap, the latter can be masked in the precipitation region by the former one. A broadening of precipitation regions (cusps) toward the equator due to an increase of the magnetic reconnection cannot be excluded: in fact, the secondary peak of the Na intensity in the equatorial belt well visible in Figure 1*b* (and still present in Figure 3*b*) could be, at least partly, driven by the IMF, and not caused by the Na migration itself. On the other hand, the increase of the magnetic reconnection can hardly explain the simultaneous poleward movement of the peak positions that may have occurred during the fourth scan - signature of major shielding of the equatorial regions - together with the increase of the equatorial emission. Thus, the previous reconnection regime should be invoked to explain the equatorial gradual enhancement.

Finally, both the proposed reconnection regimes can partly explain the observed evidences, and both contribute to depict a reliable scenario of what really occurred at Mercury and may have caused the observed dynamical behavior of the Na exospheric component.

## 5 - SUMMARY AND CONCLUSIONS

In summary, thanks to the THEMIS solar telescope in Tenerife, high resolution observations of the Na sodium exosphere of Mercury can be performed and intensity variations on time-scale of 1 hour can be followed. Models of the exospheric dynamics by Mura et al. (2009) and of the solar wind and IMF interconnections with the magnetic field of Mercury by Massetti et al. (2007) are applied to explain the observations.

Our analysis evidences the following points:

- Before allowing comparison among the data, particular care must be devoted to seeing smearing and to TAA dependence. Here we propose a procedure to avoid signal misinterpretations and to distinguish between the bulk exospheric emission and the mid latitudes intensifications.

- Cusp related peaks are observed on July 13$^{th}$ 2008, more evident in the southern region and with a minor peak in the northern hemisphere. The cusp footprints of the first scan could be modeled by means of IMF components equal to ( -10, 0, -30 nT). At the fourth scan a global enhancement of the exosphere occurs with a likely simultaneous poleward movement of the peak positions.

- In the frame of a pulsed magnetic reconnection (i.e. per-event based), the simulation of the first three scans with the exospheric dynamic model is in reasonable agreement with observed evolution. Further analyses of the other observations of the day clearly indicate that differences occur in the intensity evolution, suggesting that a new plasma

- injection event superposes, probably driven by different IMF conditions, and ifts effects on the Na exosphere are measured at about 10.50 UT (fouth scan).
- Still in the frame of a pulsed magnetic reconnection during the first scan, the observed equatorial gradual enhancement appears to bein good agreement with the Na migration scenario triggered by the plasma mid-latitudes precipitation.
- In the frame of a quasi-steady reconnection, instead, the temporal evolution of the Na emission can be modulated by the magnetic reconnection topology and reconnection rate, which can mask the features produced by the Na decay rate.
- Both the proposed reconnection regimes can partly explain the observed features, and both contribute to depict a reliable scenario.

Finally, a comprehensive explanation of the analyzed observations in July 13$^{th}$ 2008 is still not possible because of the lack of knowledge of IMF data and planetary magnetic field at the time of observation in 2008.

Nevertheless, the forthcoming study of the wide THEMIS database will increase the statistics and the knowledge of in-situ solar wind and IMF data will help understanding which process (or combination of processes) dominates in the Hermean environment. In this frame, the synergy of the Earth-based observations with the MESSENGER (Solomon et al. 2001) (now orbiting around Mercury) and BepiColombo spacecrafts (Benkhoff et al., 2010) (to be launched in 2015) is mandatory for an improved comprehension of the Hermean exosphere and its dynamics.

**TABLES AND FIGURES**

Table 1: Details of the scan sequence on July 13$^{th}$, 2008. The discrepancies in the seeing values with respect of Leblanc et al. (2009) lie within the related uncertainties.

Figure 1. Sequence of the scans obtained in July 13$^{th}$ 2008 (intensity emission, in kiloRayleigh) after preliminary reduction, including bias and sky background subtraction, as well as spectral and flux calibrations. The X-Z plane is the projection plane, with the Z axis pointing Northward; the Y axis is along the direction Earth-Mercury (with the center being the sub-Earth point). A lower cut in intensity is performed at 1000 kR.

Figure 2. Average values of Na emission from the disk of Mercury in the period July 12$^{th}$ -18$^{th}$ 2008 (crosses plus dotted line), where the vertical dotted lines separate the days of observation. The superposed fit (solid line) puts in evidence a well-defined decreasing trend during the whole period. July 13$^{th}$ (second day) is evidenced in the circle.

Figure 3. The time sequence of the Hermean sodium exosphere (from *a* to *f*) after correction from the different seeing values and taking into account also the different positions in the intensity-TAA plot. Intensity is now normalized to the 2$^{nd}$-degree fit plotted in Figure 2.

Figure 4. Simulation sequence (from *a* to *f*) as derived from the exospheric model.

Figure 5. *a*) Sketch of the magnetic field line geometry derived after updating the Massetti et al. (2007) model with the new findings from Messenger (Anderson et al., 2011). Open/closed field lines are plotted red/blue. *b*) Modeled cusp magnetic field lines (red) superimposed to the Na emission recorded at 06:52 UT.

Figure 6. Emission intensity versus time for the three regions: 'N', 'E' and 'S', identified by Z > 0.34 RM, -0.34 RM < Z <0.34 RM and Z< -0.34 RM, respectively.

Figure 7. Top: Analysis of the data emission latitude profiles averaged along the x axis. Different colors are for the six different scans of the day. The black solid profile superposed is for the 'bulk exosphere' whose maximum is normalized to the equatorial value of the first observation. The time evolution is evidenced with arrows. The y-values are the same as in Figure 3. Middle: the temporal evolution of the six scans along latitude after the subtraction of the 'bulk exosphere', normalized to 1. Bottom: the same temporal evolution after the subtraction of the 'bulk exosphere' in color code (time is from bottom to top). The numbers on the left are the time of beginning and end of each scan. The dark regions are for not-scanned periods while vertical lines are for evidencing the equator (Z = 0) and the southern and northern edges of the planet (Z = -1 and Z = 1, respectively).

**Figure1**
**Click here to download high resolution image**

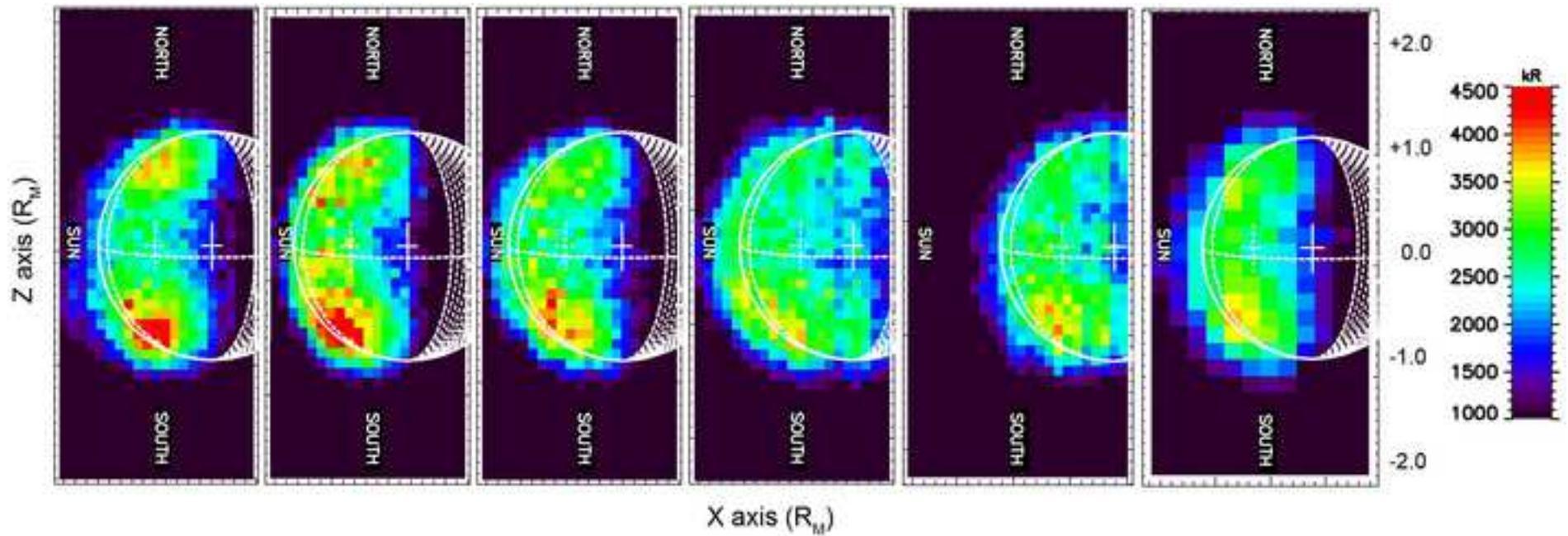

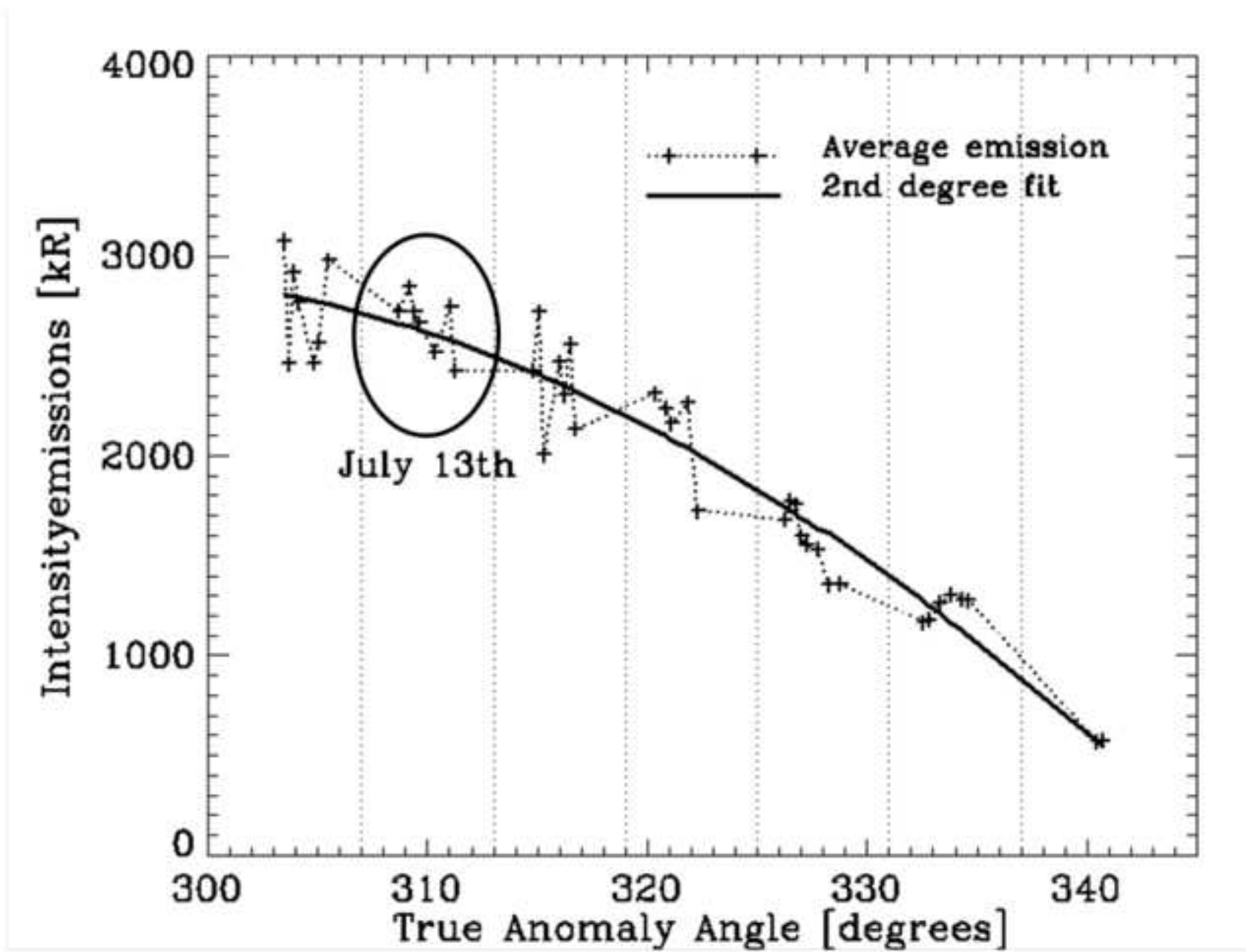

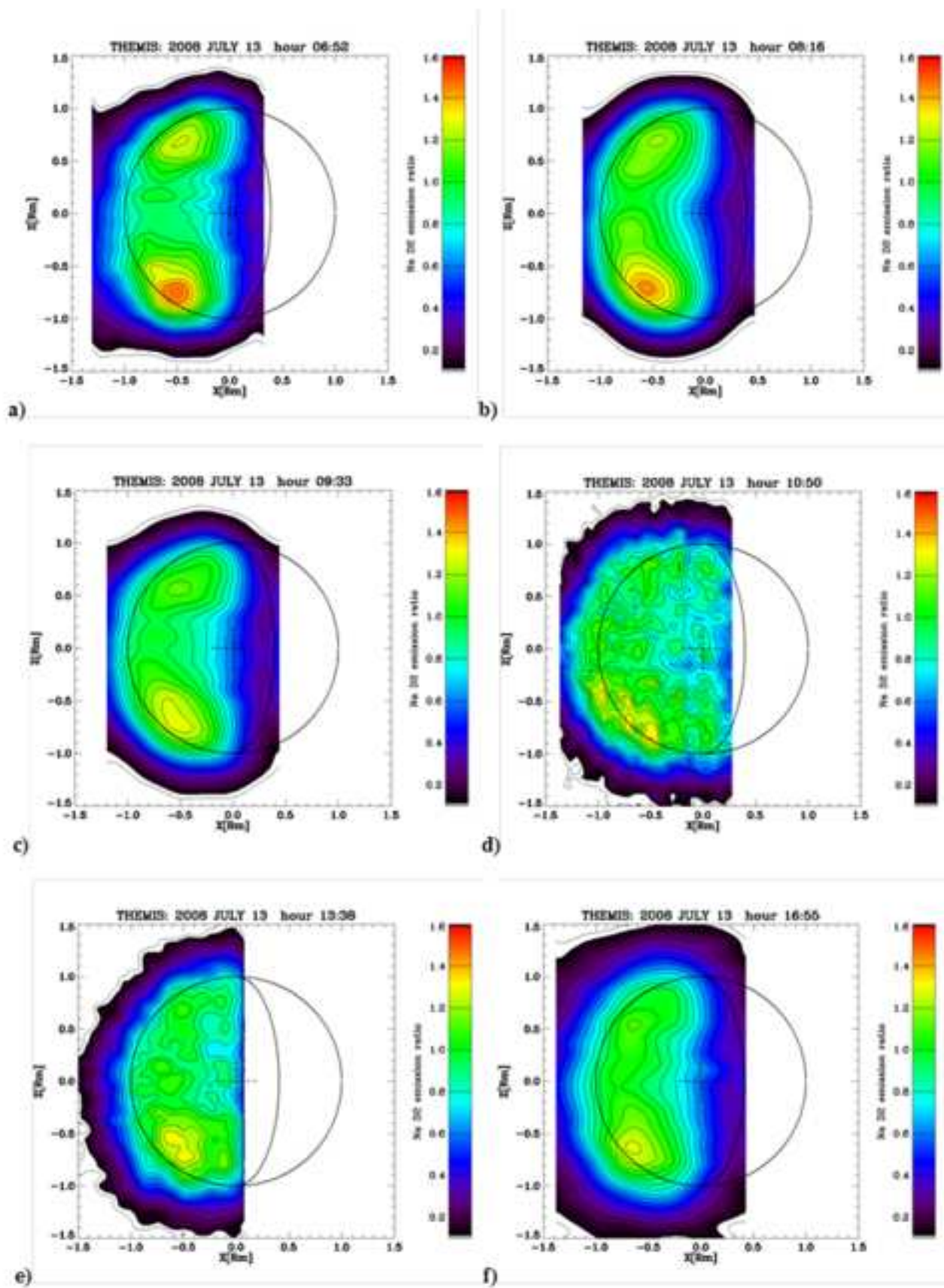

**Figure4**
**Click here to download high resolution image**

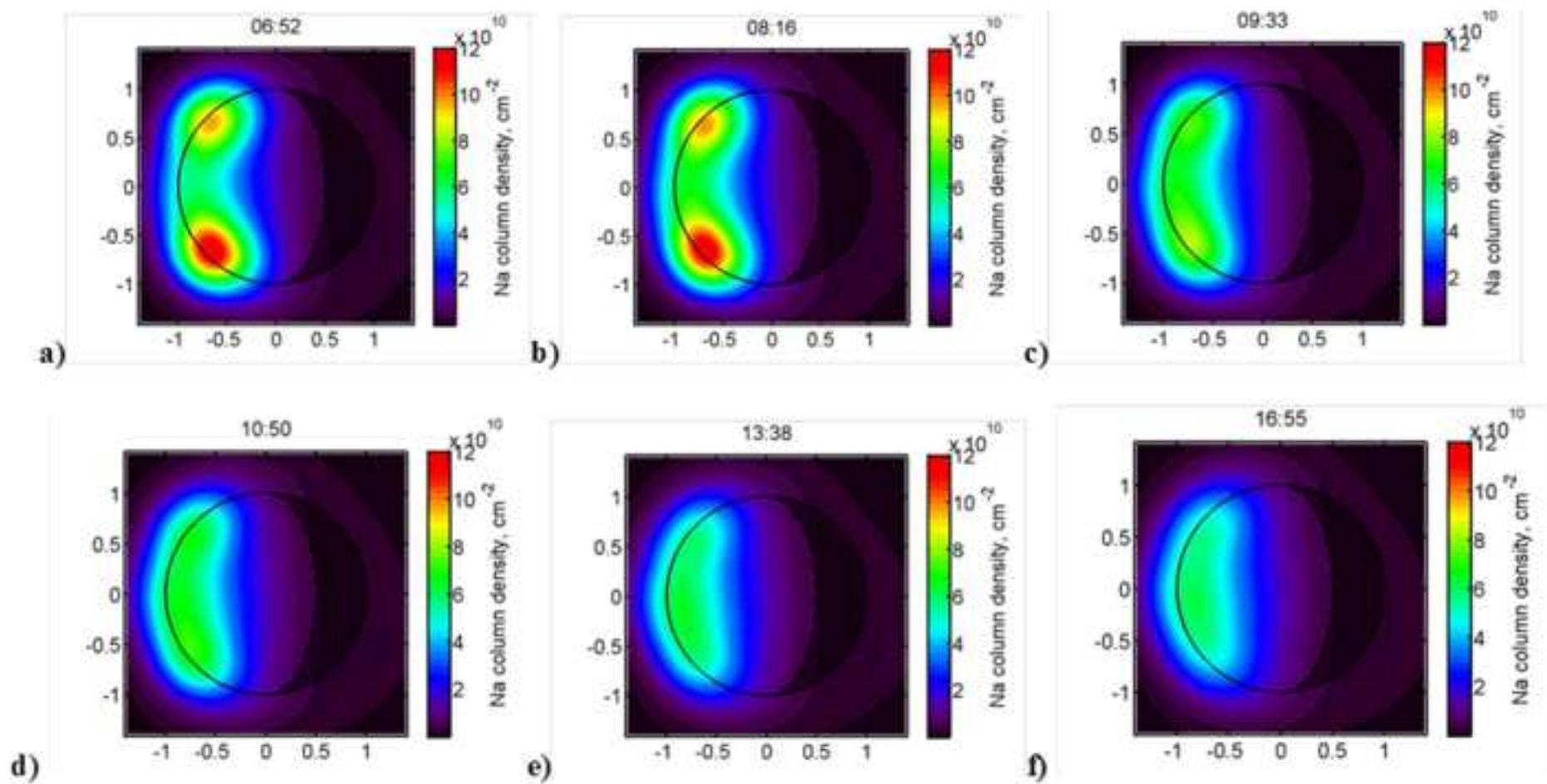

**Figure5**
**Click here to download high resolution image**

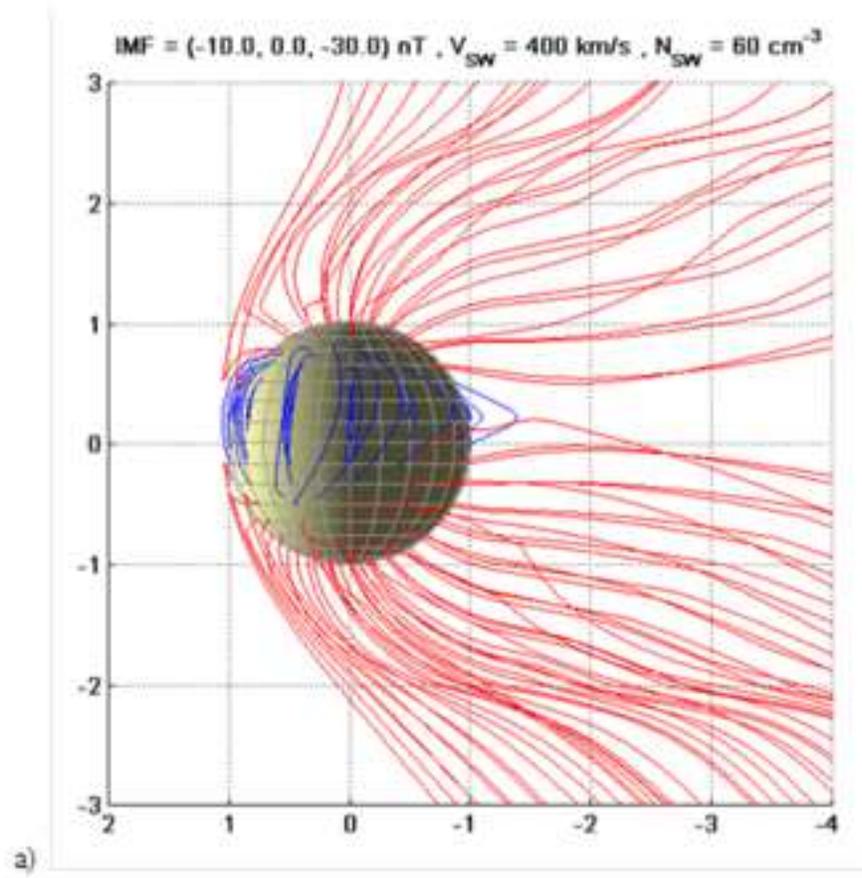

a)

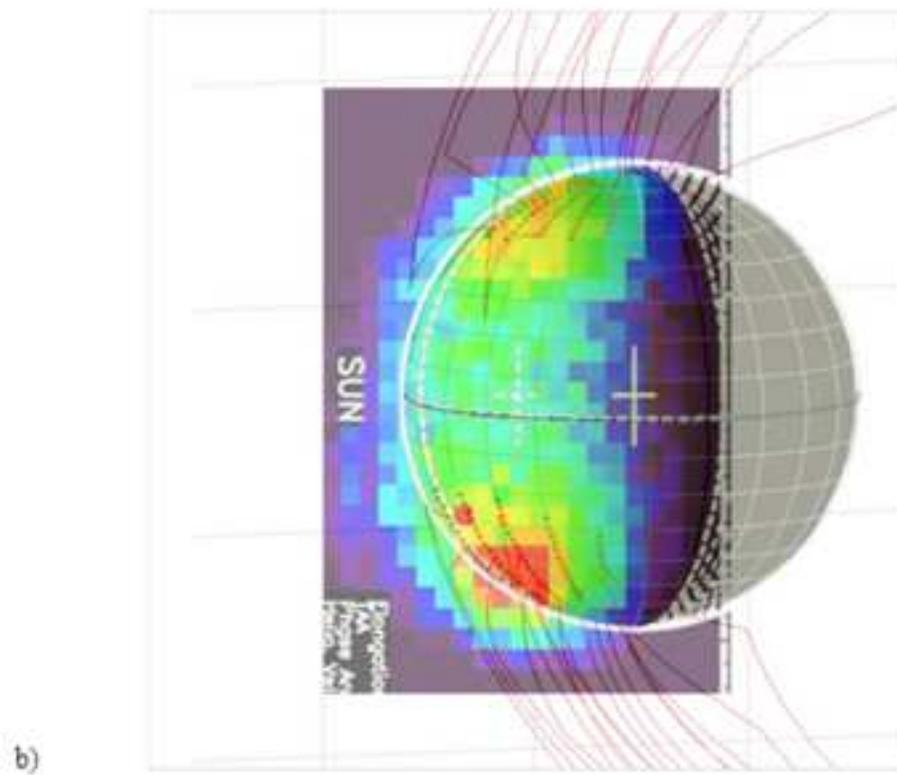

b)

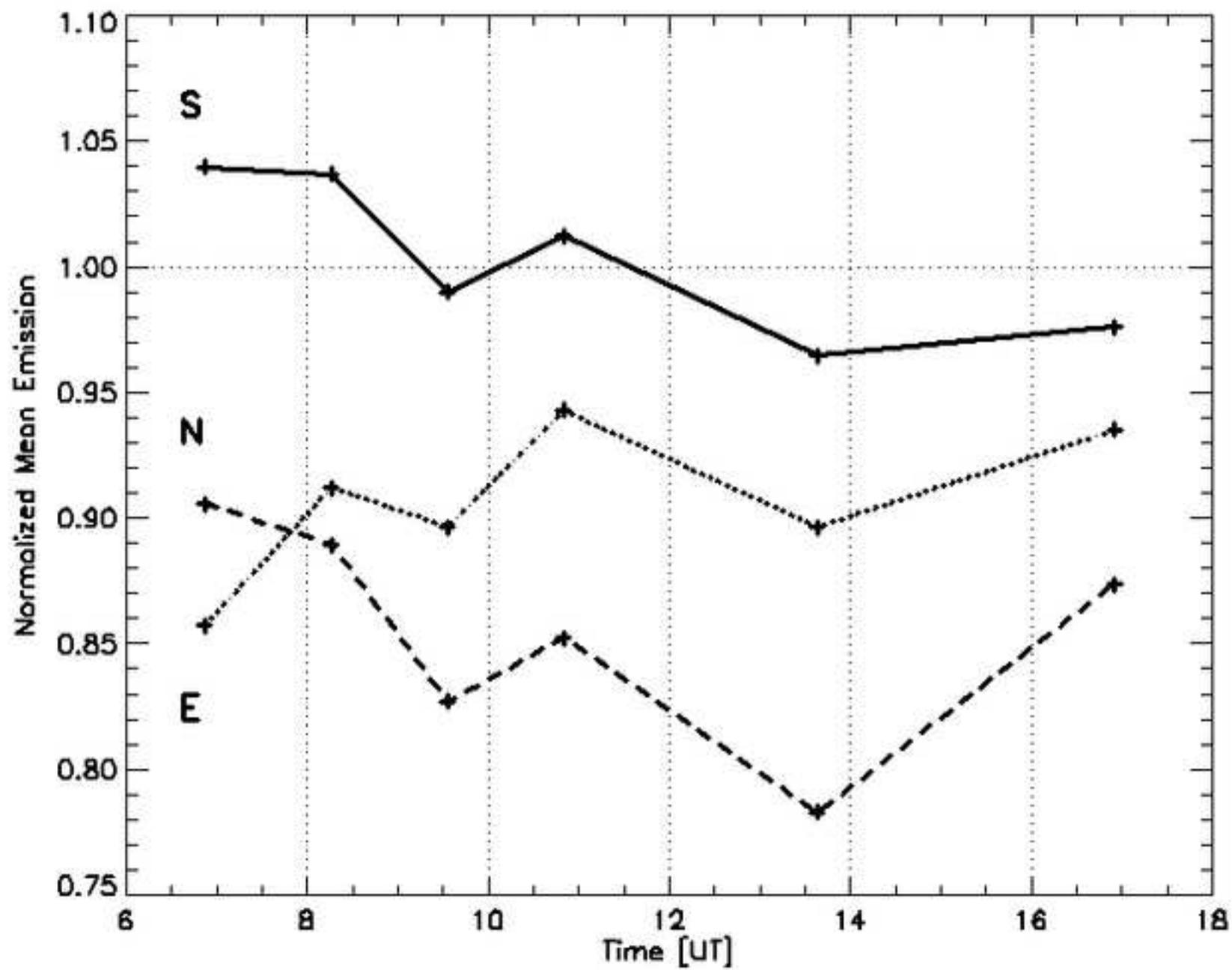

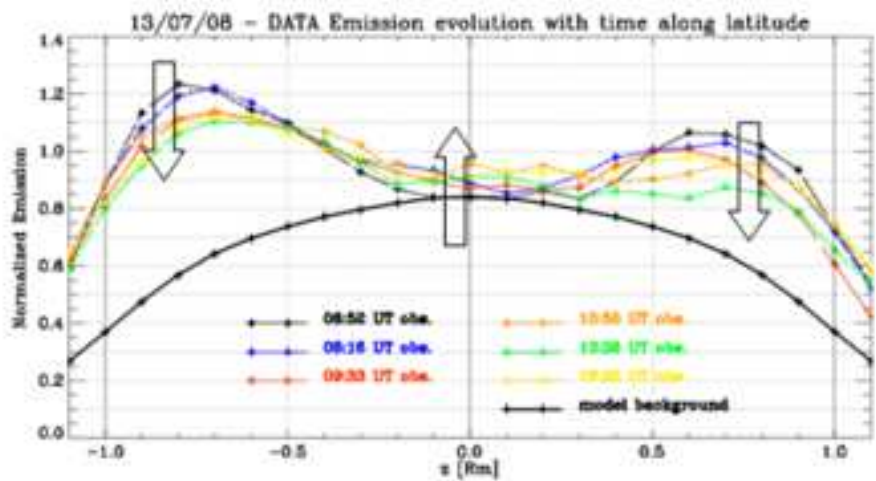
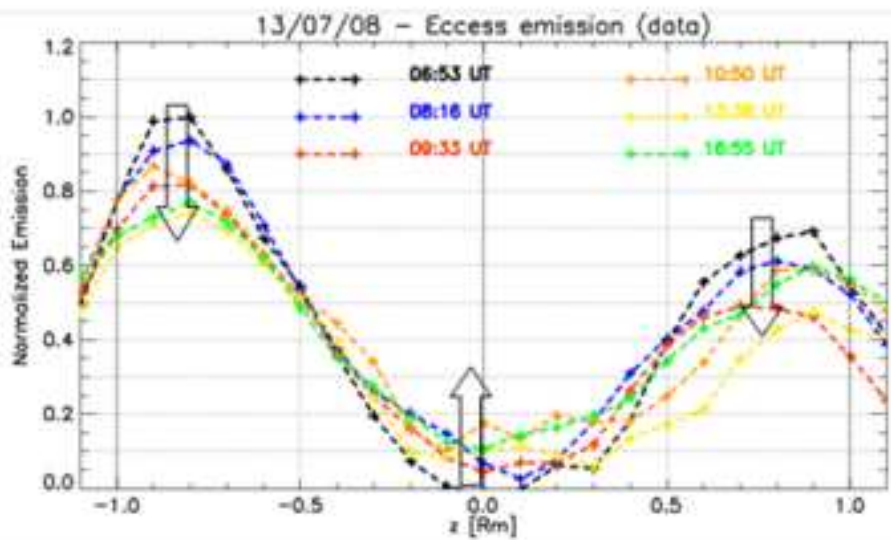
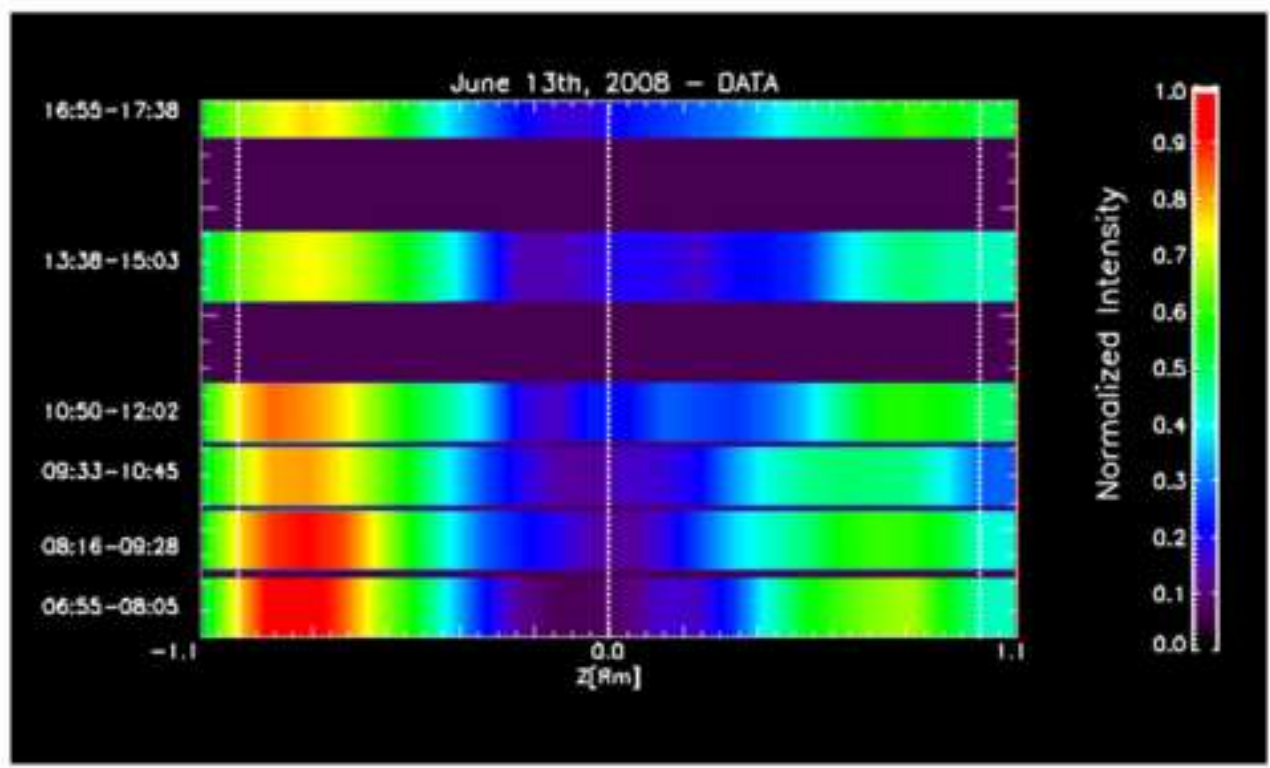

| *Scan nr.* | *Time (UT, LT+2h)* | *Resolution* | *Seeing (")* |
|---|---|---|---|
| *4* | 06:52-08:05 | high | 1.61±0.69 |
| *8* | 08:16-09:28 | high | 1.33±0.61 |
| *10* | 09:33-10:45 | high | 1.45±0.60 |
| *12* | 10:50-12:02 | high | 1.77±0.59 |
| *16* | 13:38-15:03 | high | 1.70±0.59 |
| *21* | 16:55-17:38 | low | 1.55±1.15 |

**TABLE 1**